# Radial Evolution of Thermal and Suprathermal Electron Populations in the Slow Solar Wind from 0.13 to 0.5 au: Parker Solar Probe Observations

Joel B. Abraham[1], Christopher J. Owen[1], Daniel Verscharen[1,2], Mayur Bakrania[1], David Stansby[1], Robert T. Wicks[3], Georgios Nicolaou[4], Phyllis L. Whittlesey[5], Jeffersson A. Agudelo Rueda[1], Seong-Yeop Jeong[1], and Laura Berčič[1]
[1] Mullard Space Science Laboratory, University College London, Holmbury St. Mary, Dorking RH5 6NT, UK; joel.abraham.19@ucl.ac.uk
[2] Space Science Center, University of New Hampshire, Durham, NH 03824, USA
[3] Department of Mathematics, Physics and Electrical Engineering, Northumbria University, Newcastle upon Tyne NE1 8ST, UK
[4] Southwest Research Institute, San Antonio, TX 78238, USA
[5] Space Sciences Laboratory, University of California, Berkeley, CA 94720, USA
Received 2021 June 11; revised 2022 March 28; accepted 2022 April 9; published 2022 June 1

## Abstract

We develop and apply a bespoke fitting routine to a large volume of solar wind electron distribution data measured by Parker Solar Probe over its first five orbits, covering radial distances from 0.13 to 0.5 au. We characterize the radial evolution of the electron core, halo, and strahl populations in the slow solar wind during these orbits. The fractional densities of these three electron populations provide evidence for the growth of the combined suprathermal halo and strahl populations from 0.13 to 0.17 au. Moreover, the growth in the halo population is not matched by a decrease in the strahl population at these distances, as has been reported for previous observations at distances greater than 0.3 au. We also find that the halo is negligible at small heliocentric distances. The fractional strahl density remains relatively constant at ∼1% below 0.2 au, suggesting that the rise in the relative halo density is not solely due to the transfer of strahl electrons into the halo.

*Unified Astronomy Thesaurus concepts:* The Sun (1693); Heliosphere (711); Plasma physics (2089); Solar wind (1534)

## 1. Introduction

The solar wind is a highly ionized plasma consisting of protons, α-particles, trace amounts of heavier ions, and electrons flowing continuously out of the corona and filling the heliosphere. The ions contribute to most of the mass and momentum fluxes in the solar wind due to their greater mass, while the relatively light electrons play a key role in solar wind dynamics as the main carrier of heat flux due to their much larger thermal speeds (Marsch 2006). In collisional plasmas, Coulomb collisions maintain local thermodynamic equilibrium (Feldman et al. 1975). However, the solar wind is mostly collisionless, which means that, above a certain energy, the particle velocity distribution function (VDF) can deviate from that of a classical isotropic Maxwellian equilibrium distribution. Decades of solar wind observations at heliocentric distances greater than 0.3 au have shown that the electrons in the solar wind can often be categorized into three distinct populations: the core, the halo, and the strahl (Feldman et al. 1975; Maksimovic et al. 2005; Štverák et al. 2009). The core represents the thermal part of the overall electron distribution with energy ≲50 eV. It is usually described by a (bi-)Maxwellian distribution function at 1 au (e.g., Štverák et al. 2009). The core contains 90%–95% of the total local electron density (Maksimovic et al. 2005). The Maxwellian nature of the core is attributed to collisions. At higher energies, at which collisions are less effective, nonequilibrium structures such as beams and high-energy tails can develop and survive. The halo and strahl populations represent the electrons in the suprathermal energy range (≳50 eV). The halo exhibits a greater temperature and an enhanced high-energy tail compared to the Maxwellian core distribution. It is often characterized as a (bi-) κ distribution (e.g., Štverák et al. 2009). The core and the halo are quasi-isotropic and thus show significant particle fluxes at all pitch angles. Conversely, the strahl is usually seen as a collimated, magnetic-field-aligned beam of electrons in the suprathermal energy range, moving parallel, or antiparallel, to the local magnetic field (Gosling et al. 1987). The strahl population is more often seen in the fast wind than in the slow wind (Rosenbauer et al. 1977).

Due to their weak collisionality, suprathermal electrons preserve some of their coronal characteristics and thus convey information about their coronal source regions (Scudder & Olbert 1979; Berčič et al. 2020). Therefore, precise descriptions of the electron VDF and its evolution are fundamental to determining the processes responsible for the solar wind acceleration (Jockers 1970; Lemaire & Scherer 1971; Maksimovic et al. 1997; Zouganelis et al. 2004). For example, the exospheric theory of the solar wind assumes the electron distribution to be collisionless above the exobase. It predicts that the evolution of the electron VDF through the heliosphere is driven by velocity filtration and ambipolar diffusion created by the interplanetary electric field (Maksimovic et al. 1997). Even though this model predicts the acceleration of the solar wind, the observed nature of the electron VDF in the heliosphere shows some inconsistencies with its predictions (Maksimovic et al. 2001). However, Lie-Svendsen et al. (1997) modified the model by solving the Boltzmann equation with the Fokker–Planck approximation for collisions and were able to produce results showing a strahl population consistent with that observed at 0.3 au, but with no halo present.

On average, the evolution of the core density, $n_c$, with radial distance, $r$, is in excellent agreement with expectations for an isotropically expanding gas, for which $n_c \propto r^{-2}$. In contrast, the halo and strahl populations show more complex density profiles than a steady radial expansion from 0.3 to 4 au (Maksimovic et al. 2005; Štverák et al. 2009). Under purely adiabatic conditions, the strahl population would continue to narrow in pitch angle as it propagates radially away from the Sun into regions of lower magnetic field strength, due to conservation of the magnetic moment. However, this is not generally observed, and the strahl appears to undergo significant pitch-angle scattering, as its width gradually increases with radial distance (Hammond et al. 1996; Anderson et al. 2012; Graham et al. 2017). In a simple model, Owens et al. (2008) examine the combined effects of adiabatic focusing and a constant rate of scattering on the electron populations. According to this model, a constant scattering rate dominates over the adiabatic focusing beyond ∼0.1 au, and the strahl pitch-angle width thus increases with heliocentric distance. Moreover, the strahl parallel temperature does not vary with radial distance close to the Sun (Berčič et al. 2020), which supports the assumption that the strahl carries information about the coronal temperature. However, the exact physics of the origin of the strahl is still unclear.

The origin of the radial evolution of the halo parameters remains elusive, although beam instabilities and resonant wave–particle interactions are potential mechanisms for the scattering of strahl electrons into the halo, while leaving the core relatively unaffected (Vocks et al. 2005; Saito & Gary 2007). Alternatively, Coulomb collisions (Horaites et al. 2017) or background turbulence (Saito & Gary 2007) can play similar roles in the evolution of the halo.

The solar wind near the Sun is more pristine, or less processed by transport-related effects, which means that the electron distribution function is likely to be closer to the original distribution in the outer corona of the Sun. Comparing electron distributions at different distances from the Sun with those recorded very close to the Sun enables us to improve our understanding of processes that facilitate solar wind acceleration and heating. At the same time, it allows us to probe the mechanisms that modify the distribution as the solar wind travels to greater heliocentric distances. We present the evolution of macroscopic quantities such as the density and temperature of the thermal and suprathermal populations at heliocentric distances below 0.3 au, which has not been examined using data from missions launched prior to Parker Solar Probe (PSP).

In this paper, we develop a fitting routine, which in part uses machine learning, to fit the electron VDFs measured by NASA's PSP to model distributions for the core, halo, and strahl during PSP's near-Sun encounters 2 through 5. Building on similar work by Maksimovic et al. (2005), Štverák et al. (2009), and Halekas et al. (2020), we extend the observational range to cover the region from ∼0.13 to 0.5 au, and use data with higher time resolution from PSP, to further examine the nature and evolution of the three electron populations. In Section 2, we discuss our preparation of the PSP data, and in Section 3, we describe our fitting routine and the machine learning algorithm to determine breakpoints in the distributions. Our results are presented in Section 4. We then discuss the results in Section 5 in the context of previous measurements at greater heliocentric distances. We finally provide a summary and conclusions in Section 6.

## 2. Data Handling

PSP was launched in 2018 August and will eventually achieve a closest perihelion distance of 9.86 solar radii ($R_s$) in 2024, giving us unprecedented measurements of the Sun's corona. Our analysis addresses observations over four perihelion passes or "encounters": encounter 2 (2019 March 30–April 10), encounter 3 (2019 August 16–September 20), encounter 4 (2020 January 24–February 4), and encounter 5 (2020 May 20–June 15). During the data intervals used in this study, PSP's closest perihelion is at a heliocentric distance of 0.13 au (27 $R_s$).

For the main part of our analysis, we use data from the Solar Wind Electrons, Alphas and Protons (SWEAP, Kasper et al. 2016) instrument suite. SWEAP measures the 3D electron VDF with the Solar Probe Analyzer—Electron (SPAN-E) sensor consisting of two top-hat electrostatic analyzers (ESAs): SPAN-A and SPAN-B. Together, the two ESAs measure electrons arriving from across almost the full sky using orthogonally positioned 120° × 240° fields of view (FOVs), over an energy range from 2 to 1793 eV during our measurement intervals. SPAN-A is located on the anti-ram side of the spacecraft and SPAN-B is located on the ram side. Each ESA samples over 16 azimuth, 8 elevation, and 32 energy bins. The azimuth resolution of each sensor is either 6° or 24° depending on the look direction, and covers a total of 240°. The elevation has a resolution of ∼20°. SPAN-A and SPAN-B each contain a mechanical attenuator system, which consists of a series of slits that are engaged when the particle counts approach the saturation limits of the sensor. During periods of attenuation, the total particle flux is reduced by a factor of 10. SPAN-E electron VDF measurements during encounters typically have a measurement cadence of 13.98 s. More details about the operational modes of SPAN-E are described by Whittlesey et al. (2020).

In this work, we use SPAN-E level 3 pitch-angle data. The level 3 data are provided in 32 energy bins and in 12 pitch-angle bins of width 15° with bin centers ranging from 7°.5 to 172°.5. In the production of the level 3 data set, the measurements from both sensors (SPAN-A and SPAN-B) are resampled from their intrinsic resolution onto this pitch-angle grid. These level 3 data are provided in units of differential energy flux ($cm^{-2}\,s^{-1}\,sr^{-1}\,eV^{-1}$ eV).

In order to distinguish between solar wind streams with different bulk speeds, we use data from SWEAP's Solar Probe Cup (SPC) sensor and SPAN-i. We utilize the SPC sensor to obtain the proton bulk speed moment denoted as `wp_moment` for encounters 2 and 3 (Case et al. 2020). The SPC is a Faraday cup that is mounted near the spacecraft heat shield. The SPC measurement cadence is higher than SPAN-E's, and thus in this work, our SPC moments are averaged over the SPAN-E integration times. We also use the proton bulk speed values for encounters 4 and 5 using fits to the proton measurements from SPAN-i.

We perform bi-Maxwellian fits to the proton core distribution function from the `spi_sf00_8dx32ex8a` data product, observed by SPAN-i, using the methodology described by Woodham et al. (2020), based on earlier routines developed by Stansby et al. (2018). Only the proton core speed is used from these fits in this work. The SPAN-i measurement cadence is higher than SPAN-E's and thus in this work the values are averaged over the SPAN-E integration times.

Parts of the distribution are missing due to spacecraft obstruction. To mitigate this, we remove any VDFs for which more than 20% of the data are missing. The level 3 data are converted from differential energy flux to the phase space density through

$$f(v_\parallel, v_\perp) = \frac{m_e}{V^2} J(E, \alpha) dA \, d\Omega \, dE \, dt, \quad (1)$$

where $f$ is the phase space density, $V$ is the velocity, $J$ is the differential energy flux (DEF), $\Omega$ is the solid angle, and $dt$ is the acquisition time per elevation and energy bin.

As PSP approaches the Sun, the UV radiation reaching the spacecraft surface generates increasing numbers of secondary electrons, which affect the lower energy bins. Halekas et al. (2020) account for these lower-energy secondary electrons in their fitting model by assuming the secondary electrons have a Maxwellian distribution with a fixed temperature of 3.5 eV. As our data set spans over two years we note large variations in the nature of the secondary electrons. In our fitting procedure, to avoid the effects of secondary electrons especially during the encounters, we thus ignore all data points associated with energies below 30 eV. This selection criterion makes core-temperature measurements below 30 eV less reliable than measurements at higher core temperatures.

## 3. Distribution Fitting

The fitting technique is widely used in solar and space plasma physics in order to derive plasma bulk parameters from observations (Maksimovic et al. 2005; Štverák et al. 2009; Stansby et al. 2018; Berčič et al. 2020; Halekas et al. 2020; Nicolaou et al. 2020). To capture the properties of the electrons, we analytically describe the anticipated distribution function and then fit to the measured data. Once fitted, we obtain parameters such as density, temperature, and bulk speed of each modeled population. Similar to Maksimovic et al. (2005), we fit the core electrons with a bi-Maxwellian function in the magnetic-field-aligned frame, while we fit a bi-$\kappa$ function to the halo population. Once the core and halo are fitted, Maksimovic et al. (2005) subtract the resulting core–halo distribution model from the observed distribution. They integrate the remaining population in velocity space to obtain macroscopic strahl properties. Štverák et al. (2009) perform a similar fitting routine, but modify it by fitting the suprathermal components with a truncated model, such that suprathermal components are restricted to the suprathermal parts of velocity space. Both studies show that the nonthermal halo population is modeled well by a bi-$\kappa$ and the core by a bi-Maxwellian, but Štverák et al. (2009) apply a different methodology, using a truncated $\kappa$-model to represent the strahl. This shows that, closer to the Sun, the $\kappa$ index of the strahl population approaches a value of 10, which provides a distribution that is close to a Maxwellian. In our fit model, we employ machine learning to determine the breakpoint energies of the measured distribution and then use these in the fitting routine to constrain the fits, as described in Section 3.1. The breakpoint energy is defined as the energy at which the nonthermal structures deviate from the thermal Maxwellian distribution (Feldman et al. 1975; Štverák et al. 2009). We discuss the fitting routine in Section 3.2 and the error analysis in Section 3.3.

### 3.1. Determination of Breakpoints through Machine Learning

As our fitting routine uses the breakpoint energy between the core and halo as an input, we employ the machine learning techniques described by Bakrania et al. (2020) that use unsupervised learning algorithms to determine these breakpoint energies. We also use these techniques to separate halo and strahl electrons in pitch-angle and energy space. This technique uses the $K$-means clustering method (Arthur & Vassilvitskii 2007) from the `scikit-learn` library (Pedregosa et al. 2011). $K$-means clustering works by grouping a set of observations into $K$ clusters, based on similarities between the observations. Unsupervised learning algorithms do not require the user to assign labels to training data, thereby reducing bias (Arthur & Vassilvitskii 2007). In our method, we manually set the number of clusters in the $K$-means algorithm to 2, which represents the core cluster and a suprathermal cluster. The algorithm calculates the breakpoint energy at a specific pitch angle by separating the energy distributions, at that pitch angle, into two clusters with the mid-point determined to be the breakpoint energy.

The $K$-means algorithm clusters these energy distributions by minimizing the function

$$\sum_{i=1}^{u} \sum_{j=1}^{K=2} \omega_{ij} \|x_i - \mu_j\|^2, \quad (2)$$

where

$$\mu_j = \frac{\sum_{i=1}^{u} \omega_{ij} x_i}{\sum_{i=1}^{u} \omega_{ij}}, \quad (3)$$

$$\omega_{ij} = \begin{cases} 1 & \text{if } x_i \text{ belongs to cluster } j \\ 0 & \text{otherwise,} \end{cases} \quad (4)$$

and $u$ is the number of 3-tuples at the defined pitch angle. In Equation (2), $x_i$ is defined as the vector representation of the differential energy flux tuples, where the index $i$ indicates tuples of three adjacent energy bins (i.e., energy distributions that range across three energy bins). The variable $\mu_j$ is the vector representation of two random DEF tuples, where the index $j$ labels each cluster. The $K$-means algorithm calculates the breakpoint energy by: (1) randomly selecting two DEF vectors to become the central points, or "centroids" of each cluster, $\mu_j$, (2) allocating each DEF vector, $x_i$, to its nearest centroid, by finding the smallest least-square error between that vector and the centroids, (3) determining new centroids, $\mu_j$, by averaging the DEF vectors assigned to each of the previous centroids, (4) re-allocating each DEF vector, $x_i$, to its new closest centroid, $\mu_j$, and (5) repeating steps 3 and 4 until no more new re-allocations occur. After the algorithm has computed the two clusters, the breakpoint energy at the relevant pitch angle is calculated as the center between the highest energy bin in the lower-energy cluster (which represents the core) and the lowest energy bin in the higher-energy cluster (which represents the suprathermal populations).

In order to distinguish between strahl and halo electrons, we apply this method to both pitch-angle and energy distributions. The method used for distinguishing between pitch-angle distributions is analogous to the method described above, with $x_i$ now defining a pitch-angle distribution at a certain energy. However, the $K$-means algorithm now finds the "break" in pitch angle. A detailed description of this method and an analysis of its effectiveness are provided by Bakrania et al. (2020). Arthur & Vassilvitskii (2007) details a comprehensive and more general account of the *K*-means algorithm.

After applying this method, the *K*-means algorithm outputs a list of pitch-angle bins, energy bins, and time-stamps that characterize the transition from core to suprathermal electrons. With these outputs, we obtain a set of parameters, including times when a strahl is present, strahl energies, and widths, which we use to constrain our fitting analysis.

### 3.2. Fitting of the VDF

We fit the observed distribution functions with the sum of three analytical expressions that separately describe each of the electron populations, namely the core, halo, and strahl:

$$f_e = f_c + f_h + f_s, \quad (5)$$

where $f_c$ is the fitted core, $f_h$ is the fitted halo, and $f_s$ is the fitted strahl. Following on from previous work (Maksimovic et al. 2005; Štverák et al. 2009; Berčič et al. 2020; Halekas et al. 2020), the core electrons are modeled with a two-dimensional bi-Maxwellian distribution function:

$$f_c = \frac{N_c}{\pi^{3/2} V_{\|\omega c} V_{\perp\omega c}^2} \exp\left(-\frac{V_\|^2}{V_{\|\omega c}^2} - \frac{V_\perp^2}{V_{\perp\omega c}^2}\right), \quad (6)$$

where $N_c$ is the core density, $V_{\|\omega c}$ is the core parallel thermal velocity, and $V_{\perp\omega c}$ is the core perpendicular thermal velocity. For the halo population, we fit to a bi-$\kappa$ function:

$$f_h = \frac{N_h}{V_{\|\omega h} V_{\perp\omega h}^2} \left(\frac{2}{\pi(2\kappa - 3)}\right)^{3/2} \frac{\Gamma(\kappa + 1)}{\Gamma(\kappa - \frac{1}{2})}$$
$$\times \left[1 + \frac{2}{2\kappa - 3}\left(\frac{V_\|^2}{V_{\|\omega h}^2} + \frac{V_\perp^2}{V_{\perp\omega h}^2}\right)\right]^{-(\kappa+1)}, \quad (7)$$

where $N_h$ is the halo density, $V_{\|\omega h}$ is the halo parallel thermal velocity, $V_{\perp\omega h}$ is the halo perpendicular thermal velocity, and $\kappa$ is the $\kappa$ index. For the strahl component, we use a modification to the previous works cited above and fit to a bi-Maxwellian drifting in the parallel direction at speed $U_{\|s}$ with respect to the magnetic field. Thus, the strahl is described by

$$f_s = \frac{N_s}{\pi^{3/2} V_{\|\omega s} V_{\perp\omega s}^2} \exp\left(-\frac{(V_\| - U_{\|s})^2}{V_{\|\omega s}^2} - \frac{V_\perp^2}{V_{\perp\omega s}^2}\right), \quad (8)$$

where $N_s$ is the strahl density, $V_{\|\omega s}$ is the strahl parallel thermal velocity, $V_{\perp\omega s}$ is the strahl perpendicular thermal velocity, and $U_{\|s}$ is the strahl parallel bulk velocity.

As there are 11 free parameters involved in the fit ($N_c$, $N_h$, $N_s$, $V_{\|\omega c}$, $V_{\perp\omega c}$, $V_{\|\omega h}$, $V_{\perp\omega h}$, $\kappa$, $V_{\|\omega s}$, $V_{\perp\omega s}$, $U_{\|s}$), we split our fitting process into two stages. This has the advantage of reducing the number of nonphysical fits that can arise due to the large number of degrees of freedom. The first stage is to fit only to the core + halo model and then fit to the combined model including the strahl. An example of the results of this stage is shown in the left panel of Figure 1, which presents the core and halo fits (blue and red lines respectively) to the data points from a single measured distribution (purple diamonds). In the second stage, we use the core–strahl breakpoint energy from our machine learning algorithm to constrain the relevant velocity space of the strahl electron population. This second fit captures the strahl using the drifting Maxwellian model, with the outputs of the first fit for the core and halo parameters and the strahl breakpoint energy as fixed inputs to constrain the velocity space. The right panel of Figure 1 presents the results of this strahl fit (yellow line) for the example distribution, plotted on top of the core and halo fits and the data points from the left panel. The overall fit is shown as the green trace, and from visual inspection it can be seen that a reasonable overall fit is achieved.

The fits are performed using the Levenberg–Marquardt fitting algorithm in log-space to capture the 2D electron distribution function in the field-aligned velocity space (Levenberg 1944) with each point weighted with the errors described in Section 3.3. The free parameters are constrained as follows: core, halo, and strahl density must be greater than 0; the strahl parallel bulk velocity must be less than $2.5 \times 10^7$ m s$^{-1}$; and $\kappa$ must be greater than 1.5 and less than 25.

A goodness-of-fit parameter is evaluated by comparing measured and modeled points along the perpendicular direction, because it is expected that there is no strahl present at these pitch angles, and along the parallel or antiparallel direction, which does not have the strahl (i.e., the anti-strahl direction). This allows us to capture the anisotropic nature of the core and halo populations. To evaluate the overall goodness of the fit, we evaluate the reduced $\chi^2$ parameter:

$$\chi^2 = \frac{1}{n-m} \sum_i \frac{(O_i - C_i)^2}{\sigma^2}, \quad (9)$$

where $O_i = \log(\tilde{f}_i / 1 \text{ s}^3 \text{ m}^{-6})$ are the measured data based on the measured full distribution function $\tilde{f}_i$, $C_i = \log(f_e / 1 \text{ s}^3 \text{ m}^{-6})$ are the fitted data, $n$ is the number of fitted data points, $m$ is the number of variables to fit, and $\sigma^2$ is the variance of $\log(f / 1 \text{ s}^3 \text{ m}^{-6})$.

We assume that the bulk speeds of the core and halo populations are zero in our fit models in the instrument frame. Consequently, any measured distributions with significant nonzero drifts manifest as a large reduced $\chi^2$ value and are excluded from the analysis. Once that is done, we undertake the analysis of the features of the suprathermal populations by taking partial moments of the fitted curve by integrating over the part of velocity space constrained by the breakpoints obtained from the machine learning algorithm described in Section 3.1.

### 3.3. Error Analysis

We model the overall measurement error as a combination of that given by Poisson statistics and an additional error that reflects the likely systematic error in the instrument measurement, arising due to the finite micro-channel plate efficiency and uncertainty in other instrumental effects, which we combine and capture here as an effective overall uncertainty in the instrument geometric factor. The Poisson error is the dominant error source when the number of counts is small. We quantify the relative error in the geometric factor as 10%, a value that has been adopted following direct discussions with the data provider team.

In the creation of the SWEAP VDF data, the value of the distribution function *f* at a given energy, azimuth, and elevation

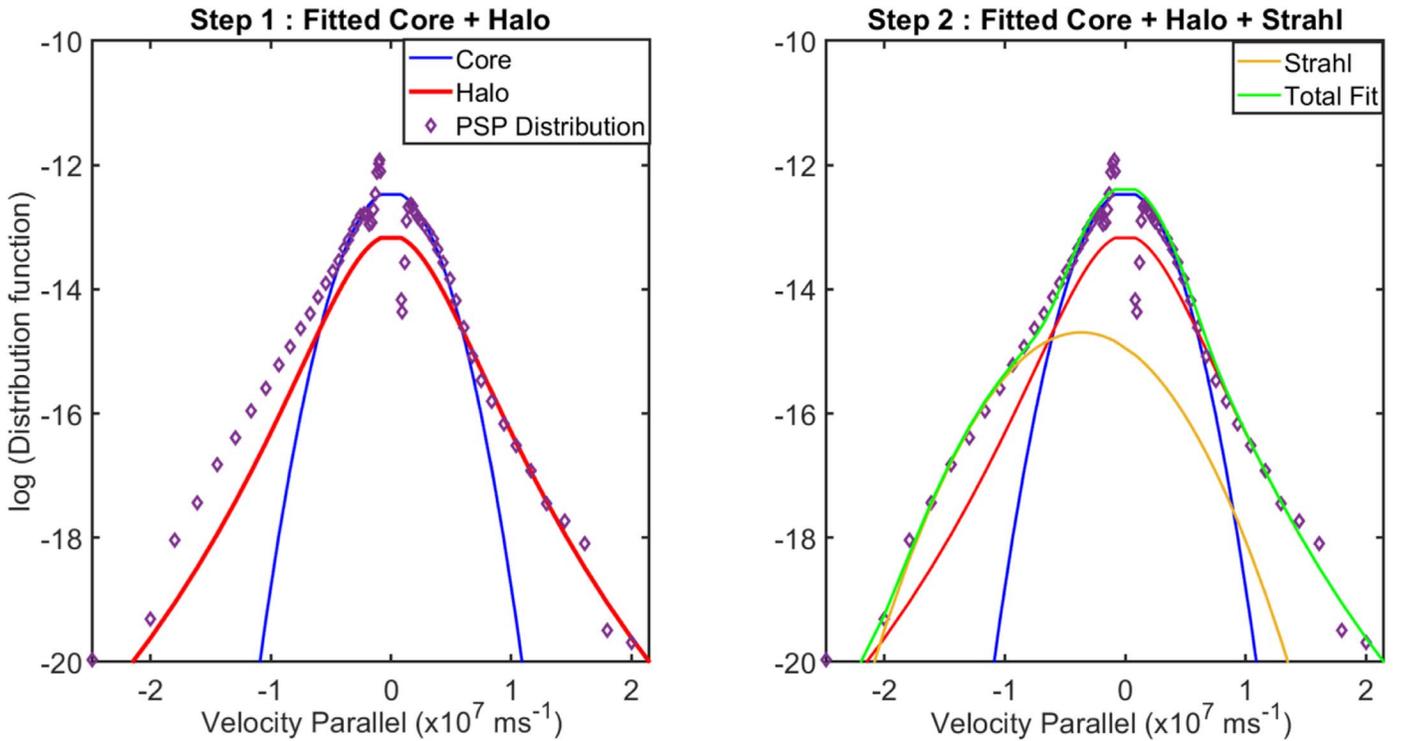

**Figure 1.** Two-step fitting process. The purple diamonds mark the measured distribution at 0.2940 au on 2019 August 25 at 03:28:28 UT, the blue curve represents the fit for the bi-Maxwellian core, and red the fit for the bi-$\kappa$ halo. The yellow curve represents the fit for the drifting bi-Maxwellian strahl. The panel on the left shows the core and halo fits for the measured distribution. The second fit is shown in the right-hand panel where the strahl is fitted. The green curve represents the total fit.

is calculated based on the raw counts $C$ as

$$f = \frac{m_e{}^2 C}{2\,\Delta t\,E^2 G}, \qquad (10)$$

where $\Delta t$ is the counter readout time, $G$ is the geometric factor, and $E$ is the energy. Based on Gaussian error propagation, the Poisson error and the uncertainty in the geometric factor lead to the following result for the variance of the data points (i.e., of $\log(f)$) in our measured distribution function:

$$\sigma^2 = \left(\frac{m_e}{\ln(10) E \sqrt{2\,\Delta t\,Gf}}\right)^2 + \left(\frac{1}{\ln(10)} \frac{\Delta G}{G}\right)^2, \qquad (11)$$

where $\Delta G/G = 0.1$ is the relative error in the geometric factor. In our analysis, we only include fits that have $\chi^2 \leqslant 1$.

Overall, we examine over 450,000 electron velocity distribution functions obtained by PSP SWEAP from the years 2019 and 2020. After applying the $\chi^2$ limit and further removal of some clearly nonphysical fits, we obtain ∼300,000 fits for further analysis, of which 220,000 have an associated measurement of solar wind speed from SPC or SPAN-i.

### 4. Results

Most of the measurements during this time period have proton bulk speeds of less than 400 km s$^{-1}$, which we classify as the slow wind. We split the data into 50 equal-width radial distance bins, and the median value of a given parameter of interest in each radial distance bin is calculated. We calculate the upper and lower error bars for each radial distance bin as the upper and lower quartiles respectively.

Figure 2 shows the radial evolution of the averaged fitted parameters. Panel (a) shows the averaged core density as a function of the heliocentric distance. This is broadly in line with the expectations for a radial isotropic expansion of this population. The $r^{-2}$ trend is represented by the solid green curve. Below 0.2 au, the halo density (panel (b)) shows only a moderate dependence on heliocentric distance within the error bars. However, over the same radial distance range, the strahl density (panel (c)) has a clearly steeper gradient, which is more significant.

The thermal speed of the fitted core distribution (panel (e)) decreases with radial distance in this range. The thermal speed of the fitted halo distribution initially increases from $2.2 \times 10^6$ m s$^{-1}$ at 0.13 au to $3.5 \times 10^6$ m s$^{-1}$ at 0.23 au, and thereafter the thermal speed decreases. The parallel thermal velocity is enhanced above the perpendicular thermal velocity at all distances shown, indicating a persistent anisotropy in the parallel direction for the halo population.

The radial evolution of the $\kappa$ value for the fitted halo distribution is shown in panel (d). The $\kappa$ parameter provides a measure of the nonthermal state of the halo population. As $\kappa$ tends to infinity, the distribution becomes closer to a Maxwellian. For the slow solar wind regime shown here, $\kappa$ is low at ∼4 for the shortest distances sampled. The $\kappa$ value rises from ∼4 to ∼12 between 0.13 and ∼0.24 au before steadily decreasing over the rest of the distance range shown.

We also observe that the fit to the strahl component shows a strong decrease in density with distance in both solar wind regimes (panel (c)). The strahl thermal speed component is $V_{\parallel ws} > V_{\perp ws}$ closer to the Sun, and slowly decreases with radial distance, such that the strahl distribution is isotropic (within error bars) by ∼0.2 au, unlike the core and halo, which both show clear declines as distance increases.

To examine the complex radial evolution of the suprathermal population, we numerically integrate the total fitted curve over velocity space using the breakpoints to define the energy and/or pitch-angle integration limits for the core, halo, and strahl populations. In Figure 3, we present the integrated density evolution of the three electron populations on a common scale with heliocentric distance. Here the integrated core density data are shown in blue, the integrated halo density in orange, and the integrated strahl density in yellow. The two suprathermal populations are at least an order of magnitude lower in density than the core population across the entire distance range shown. As mentioned above, the core density falls as $r^{-2}$ up to 0.25 au. From 0.25 au, we note a deviation of the core electron population from the $r^{-2}$ expansion curve.

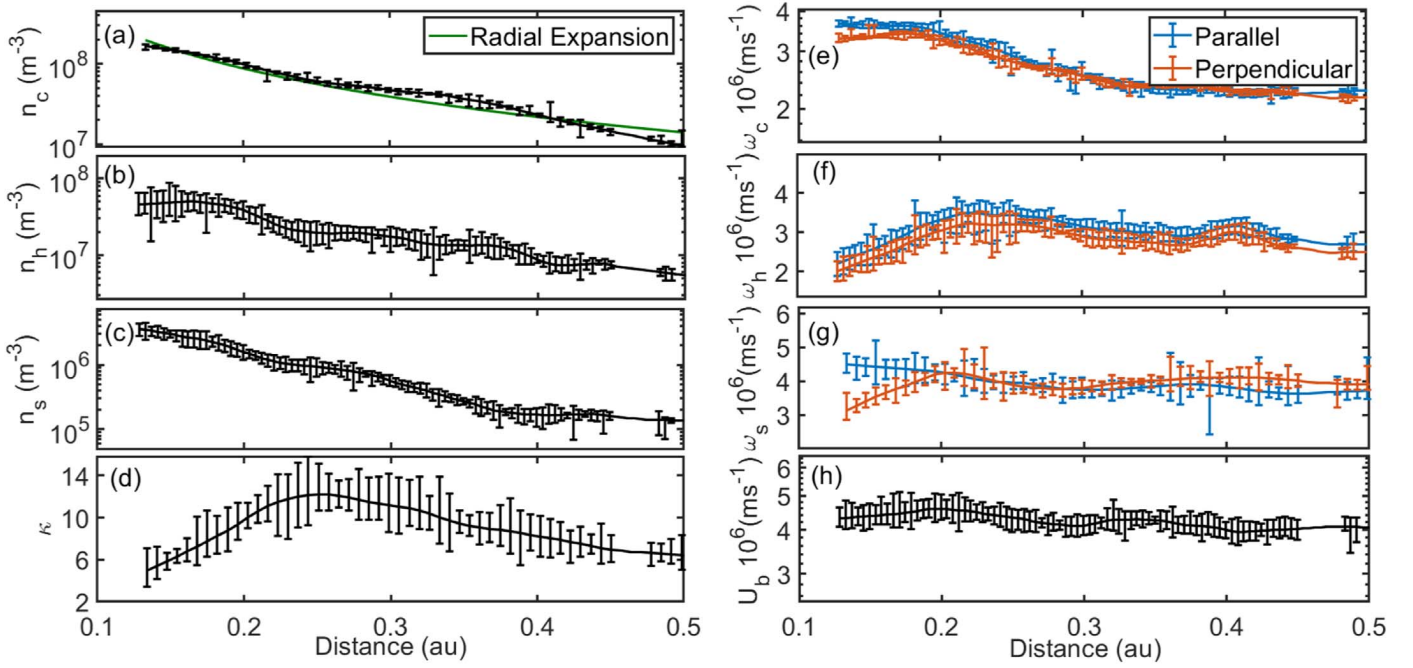

**Figure 2.** The radial evolution of the fit results for solar wind at speeds less than 400 km s$^{-1}$. Panel (a) shows the radial evolution of the core density and the black dashed curve shows the expected evolution of an isotropically expanding gas. Panels (b) and (c) represent the radial evolution of the halo and strahl populations respectively. Panel (d) shows the radial evolution of $\kappa$ for the fitted halo population. Panels (e), (f), and (g) represent parallel and perpendicular thermal speeds of the core, halo, and strahl respectively. Panel (h) shows the radial evolution of the strahl bulk speed.

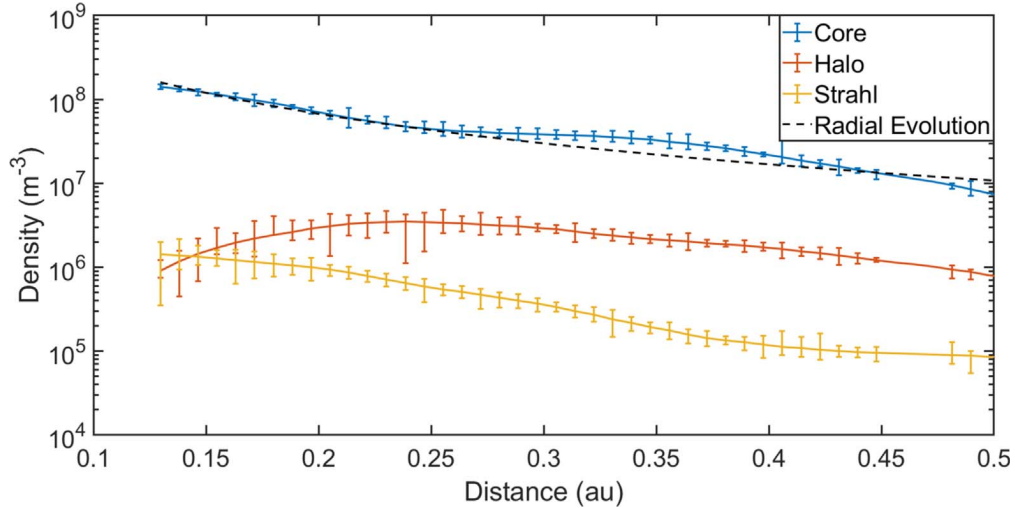

**Figure 3.** The blue curve represents the core density, the orange curve is the halo density, the yellow curve is the strahl density, and the black dashed curve is the theoretical prediction for an isotropically expanding gas.

Figure 3 shows that from 0.2 au outwards, the halo (orange curve) makes up most of the suprathermal population, while the strahl makes up most of this population below 0.2 au. Figure 3 also shows that the evolution of the suprathermal population with radial distance does not follow an $r^{-2}$ trend. Below 0.25 au, the halo density decreases with radial distance while there is a small increase in the strahl density. From 0.25 au onwards, both populations show a steady decline in density with increasing radial distance.

To remove the effects of expansion, we look at the relative densities of the three electron populations with respect to the total local electron density in a similar way to Štverák et al. (2009). Figure 4 shows the relative density of the core population is above 90% across the full distance range sampled. Thus, the combined density of the suprathermal populations (shown in these plots by the purple curve) makes up less than 10% of the total electron density observed at any distance. From 0.124 to 0.2 au, the relative halo density (orange curve) increases from less than 1% to ∼10% of the total electron density in all cases. Between 0.15 and 0.2 au, there is a point in each plot at which the relative halo density is equal to the relative strahl density, which we refer to as the *halo–strahl crossover point* in this paper. It is not straightforward to determine the exact location of the halo–strahl crossover point due to the size and overlaps in the error bars. The relative strahl density stays approximately constant (∼1%) below 0.2 au but there is a sharp rise in relative halo density from less than 1% to ∼7%. The total fractional density of the combined suprathermal populations rises from ∼1% at the closest distances sampled (∼0.13 au) to almost 10% above 0.25 au.

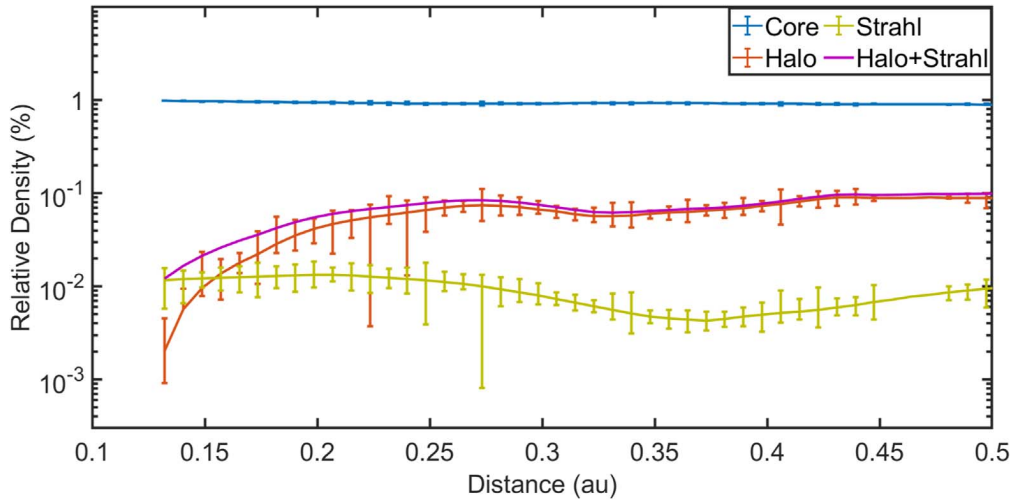

**Figure 4.** The blue curve represents the core density, the orange curve is the halo density, the green curve is the strahl density, and the purple curve is the total suprathermal population.

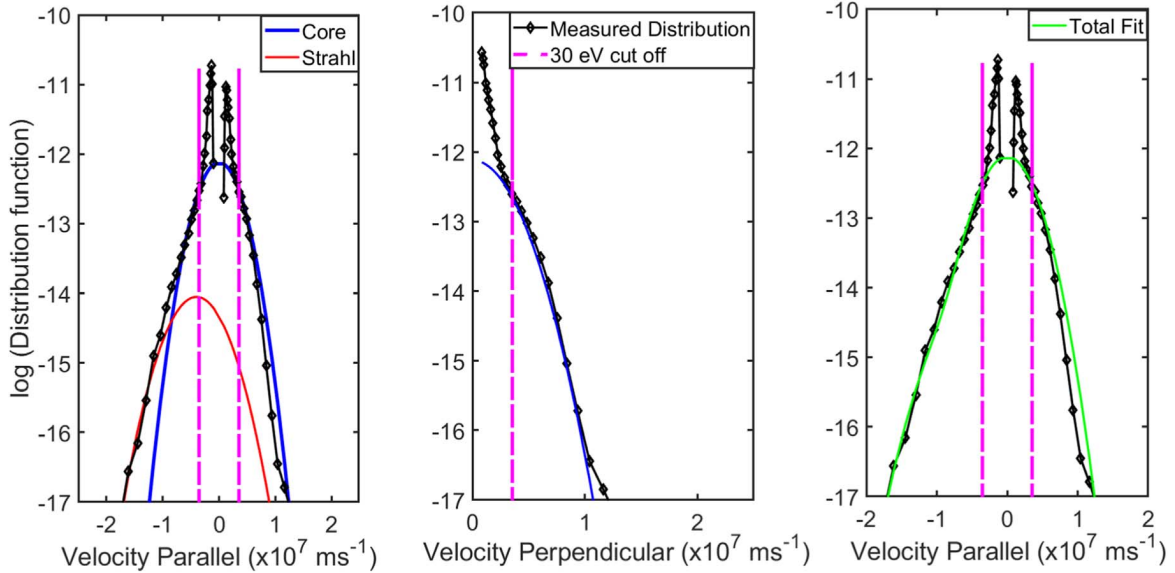

**Figure 5.** A representative distribution recorded at a distance of ∼0.13 au. The black trace with diamonds shows the measured distribution. The blue trace represents the output of our fitting routine for a Maxwellian core, and the red trace represents the output for the fit to a Maxwellian strahl drifting along the *B*-field direction. The pink vertical dashed lines represent the 30 eV measurement energy below which we do not fit to data due to secondary contamination, as discussed in the text. The left-hand panel shows a cut along the parallel velocity direction, while the middle panel shows the cut along $V_\parallel = 0$ in the perpendicular direction. In the right-hand panel, the green trace shows the final combined fitted curve to the measured distribution. This panel indicates an excellent fit to the data without the need to infer a third fit for the halo model, meaning that the halo contribution to this distribution is negligible.

To examine this in more detail, we investigate the average shape of the distribution function below the halo–strahl crossover point. For this selection of data, most of the VDFs can be described well with just the core and strahl elements of the model fit, with no explicit need to include a halo model, as seen from the example distribution/fits illustrated in Figure 5.

### 5. Discussion

The relative densities and radial trends of properties of the core electron population are broadly in agreement with previous observations by Maksimovic et al. (2005) and Štverák et al. (2009) for distances ≳0.28 au. These authors demonstrate, for this distance range, that the relative density of the strahl is greater than the relative density of the halo closer to the Sun (below ∼0.6 au). Contrary to these previous results, our observations show that the relative density of the two suprathermal electron populations does not evolve in an inverse manner below 0.3 au, where the strahl density decreases as the halo density increases. Below 0.2 au, the total fractional density of the suprathermal population is not constant, but drops from ∼10% to ∼1%. This implies that with increasing distance from the Sun, there must be some process or processes driving an increase in the total number of electrons in the suprathermal energy range. A candidate source for these electrons is the core in this region. If this is the case, then it is possible that the quasi-isotropic nature of the halo can be explained by a process that creates the halo from the core population. Extrapolating the lines in Figure 4 to distances closer to the Sun, we notice that for the slow solar wind, the distribution function is possibly composed of a core and a strahl model without a significant halo component.

The fact that the fractional strahl density remains relatively constant at ∼1% and the fractional halo density increases from less than 1% at 0.124 au to ∼7% at 0.2 au shows that the halo cannot be formed from the scattering of strahl alone as suggested by Štverák et al. (2009). Thus, there appears to be more than one process contributing to the growth of the halo population. This may be a multistage process in which, say, a fraction of the core population is accelerated to suprathermal energies by a resonant wave–particle interaction or other plasma processes. Alternatively, larger-scale dynamics may play a role, such as the field-aligned acceleration of reconnection outflow beams, followed by scattering in pitch angle to form the halo. Further analysis is needed to confirm the nature of any such coupling between the core and the suprathermal population. We also define the halo–strahl crossover point, described above as the point where the halo density and strahl density are equal. Below the halo–strahl crossover point, most of the suprathermal population moves along the field line while above this point most of the suprathermal population is present at all pitch angles. This point may be important in the study of processes that concern the evolution of the suprathermal populations. However, the fractional trends in Figure 3 show that the total suprathermal population continues to decrease in the same radial range.

We also specifically examine electron VDFs that are recorded at radial distances below the halo–strahl crossover point. Electron distributions below 0.15 au can in general be well described with only a core and a strahl model as shown by Figure 5. Another feature we often note at the closest distances sampled is a deficit in the measured distribution function, with reference to the core fit, in the anti-strahl direction. Halekas et al. (2020) examine the first two orbits and report a similar truncation in the Maxwellian. This deficit is not included explicitly by our model, and this means our core density at the closest distances to the Sun may be slightly overestimated. If the deficit is sufficiently large, then this would result in a large reduced sum of the squares and be discarded from our analysis.

Another interesting result from our analysis is the variation in the $\kappa$ value with radial distance. The variation in $\kappa$ indicates changes in the shape of the high-energy tails with radial distance. At the closest distance sampled, the $\kappa$ value is ∼3, and it rises to 6 around 0.25 au. When we compare this rise in $\kappa$ with the trend in halo density shown by Figure 4, we notice that the halo density is less than 0.01% and rises to a few percent.

From Figures 2(e)–(g), we observe the nonadiabatic nature of the slow solar wind. At the closest distance sampled, there seems to be a persistent anisotropy in all three electron populations, with the strahl exhibiting the strongest parallel anisotropy. The core population cools with radial distance but with gradients in the thermal speed. This shows that the cooling rate varies with radial distance while remaining quasi-isotropic with the thermal speeds within the error bars. However, the halo thermal speed initially rises from $2 \times 10^6$ m s$^{-1}$ at 0.13 au to $3.7 \times 10^6$ m s$^{-1}$ at 0.25 au and then decreases with radial distance. The initial rise can be attributed to the growth of the halo as more particles populate the upper halo energy range. We are unaware of any theory that explains these thermal trends. Further research into what drives these trends is needed.

As is evident from Figure 2, the strahl parallel thermal velocity does not vary with radial distance when fitted to a drifting bi-Maxwellian model. This result is also reported by Berčič et al. (2020). It is consistent with a recent kinetic model for the strahl evolution in the inner heliosphere, which also shows that the strahl parallel temperature and bulk velocity are constant with heliocentric distance (Jeong et al. 2022). In exospheric models, the strahl is believed to carry information about the exobase (Jockers 1970), which means that the constant strahl parallel temperature and bulk speed provide critical information about the coronal electrons at the wind's origin. The strahl parallel thermal speed from our fits is approximately the same magnitude as the typical temperature of the corona ($\approx 10^6$ K). Further analysis of data closer to the Sun obtained from future PSP encounters will be needed to confirm whether the strahl parallel temperature indeed preserves the coronal electron temperature. The core and strahl have a parallel anisotropy closer to the Sun at 0.13 au, but this anisotropy decreases with radial distance and approaches isotropy within the statistical errors. Another new finding we show with our work is that the strahl parallel bulk speed stays roughly constant within the error bars. This means that the strahl is potentially a useful indicator of the origins of the source regions of the solar wind.

In exospheric solar wind models, the electrons with energy less than the electric potential at a given radial distance are reflected and are trapped in a potential well (Maksimovic et al. 2001). The deficit in the distribution function that we observe in the anti-strahl direction could then be a result of this trapping boundary. However, in this theory, the cutoff in the distribution is quasi-discontinuous, while we observe a smoother drop below the Maxwellian VDF values in the sunward side of the VDF. We also observe that this signature becomes weaker with radial distance. This change in the signature may be explained by collisions, which are usually ignored by exospheric models beyond the exobase. Regardless, more research is required to better understand these deficits by quantifying the point where the Maxwellian truncates, since they potentially give insight into the role of the interplanetary electrostatic potential in the acceleration of the solar wind.

## 6. Conclusions

We apply a new fitting routine to electron VDF measurements, which for the first time incorporates breakpoint energies obtained from a machine learning algorithm (Bakrania et al. 2020). This new technique is applied to a large data set of PSP SPAN observations at varying distances from the Sun. We use our fitting results to investigate the evolution of the core, halo, and strahl for encounters 2, 3, 4, and 5. We show that the core makes up more than 90% of the total electron density for all the distances sampled, whereas the nonthermal electrons make up less than 10%, as previously observed for distances >0.3 au (Maksimovic et al. 2005; Štverák et al. 2009; Halekas et al. 2020). The radial $r^{-2}$ dependence of the core density extends below 0.3 au. We also show that the relative suprathermal population increases from ∼1% at the closest distances sampled to ∼10% around 0.22 au, which indicates that there is a relative increase in the nonthermal particle densities over the inner regions of the heliosphere.

Our analysis does not reveal a distinct inverse relationship between the halo and strahl populations below 0.25 au. Rather, we find that the relative strahl density stays approximately constant while the relative halo density increases. We introduce a point called the halo–strahl crossover point, where the relative halo density is equal to the relative strahl density. At the closest distances sampled below this point, the distribution can generally be well modeled with only a core and strahl model with little/no contribution from the halo model. The low halo density closer to the Sun suggests the halo is diffused and drops below the one-count sensitivity level of the instrument. Another key feature we report is that below the halo–strahl crossover point, we generally see a distinct deficit in the core population in the anti-strahl direction. This indicates that there are fewer particles in the part of velocity space corresponding to particles returning in the direction of the Sun than expected from the Maxwellian fit. Such a cutoff in the distribution is predicted by Maksimovic et al. (2001). However, above the halo–strahl crossover point, we do not generally see such a deficit in the distribution with respect to the modeled fits.

In the future, we aim to quantify the nature of these deficits with a bespoke fitting routine that helps us to better understand the role of the interplanetary electrostatic potential in solar wind acceleration. We also aim to examine solar wind energetics to understand the mechanisms at play that lead to the growth of the nonthermal populations. With the advent of Solar Orbiter, an interesting avenue for further research is to look at alignments with PSP to study the same plasma parcel with this technique.

We acknowledge the NASA Parker Solar Probe Mission and SWEAP team led by Justin Kasper for use of data. The authors are grateful to PSP instrument teams for producing and making the data used in this study publicly available. The SPAN-E level 3 data are obtained from http://sweap.cfa.harvard.edu/pub/data/sci/sweap/spe/L3/spe_sf0_pad/. J.B.A. is supported by the Science Technology and Facilities Council (STFC) Studentship ST/T506485/1. D.V. is supported by STFC Ernest Rutherford Fellowship ST/P003826/1. M.R.B. is supported by a UCL Impact Studentship, jointly funded by the ESA NPI contract 4000125082/18/NL/MH/ic. C.J.O., D.V., D.S. and L.B. receive support under STFC grant ST/S000240/1. R.T.W. is supported by STFC Consolidated Grant ST/V006320/1. J.A. A.R. is supported by ESA NPI contract 4000127929/19/NL/MH/mg and ICETEX, reference 3933061. P.L.W. acknowledges SWEAP support under NASA PSP Phase E contract NNN06AA01C. This work was discussed at the "Joint Electron Project" at MSSL.

## ORCID iDs

Joel B. Abraham https://orcid.org/0000-0002-6305-3252
Christopher J. Owen https://orcid.org/0000-0002-5982-4667
Daniel Verscharen https://orcid.org/0000-0002-0497-1096
Mayur Bakrania https://orcid.org/0000-0001-6225-9163
David Stansby https://orcid.org/0000-0002-1365-1908
Robert T. Wicks https://orcid.org/0000-0002-0622-5302
Phyllis L. Whittlesey https://orcid.org/0000-0002-7287-5098
Jeffersson A. Agudelo Rueda https://orcid.org/0000-0001-5045-0323
Seong-Yeop Jeong https://orcid.org/0000-0001-8529-3217
Laura Berčič https://orcid.org/0000-0002-6075-1813

## References

Anderson, B. R., Skoug, R. M., Steinberg, J. T., & McComas, D. J. 2012, JGRA, 117, A04107
Arthur, D., & Vassilvitskii, S. 2007, in Proc. 18th Annual ACM–SIAM Symp. on Discrete Algorithms (New Orleans, LA: Society for Industrial and Applied Mathematics), 1027
Bakrania, M. R., Rae, I. J., Walsh, A. P., et al. 2020, A&A, 639, A46
Berčič, L., Larson, D., Whittlesey, P., et al. 2020, ApJ, 892, 88
Case, A. W., Kasper, J. C., Stevens, M. L., et al. 2020, ApJS, 246, 43
Feldman, W., Asbridge, J., Bame, S., Montgomery, M., & Gary, S. 1975, JGR, 80, 4181
Gosling, J. T., Baker, D. N., Bame, S. J., et al. 1987, JGRA, 92, 8519
Graham, G. A., Rae, I. J., Owen, C. J., et al. 2017, JGRA, 122, 3858
Halekas, J. S., Whittlesey, P., Larson, D. E., et al. 2020, ApJS, 246, 22
Hammond, C. M., Feldman, W. C., McComas, D. J., Phillips, J. L., & Forsyth, R. J. 1996, A&A, 316, 350
Horaites, K., Boldyrev, S., Wilson, L. B. I., Viñas, A. F., & Merka, J. 2017, MNRAS, 474, 115
Jeong, S.-Y., Verscharen, D., Vocks, C., et al. 2022, ApJ, 927, 162
Jockers, K. 1970, A&A, 6, 219
Kasper, J. C., Abiad, R., Austin, G., et al. 2016, SSRv, 204, 131
Lemaire, J., & Scherer, M. 1971, JGR, 76, 7479
Levenberg, K. 1944, QApMa, 2, 164, http://www.jstor.org/stable/43633451
Lie-Svendsen, Ø., Hansteen, V. H., & Leer, E. 1997, JGRA, 102, 4701
Maksimovic, M., Pierrard, V., & Lemaire, J. 2001, Ap&SS, 277, 181
Maksimovic, M., Pierrard, V., & Lemaire, J. F. 1997, A&A, 324, 725
Maksimovic, M., Zouganelis, I., Chaufray, J.-Y., et al. 2005, JGR, 110, 110
Marsch, E. 2006, LRSP, 3, 1
Nicolaou, G., Wicks, R., Livadiotis, G., et al. 2020, Entrp, 22, 103
Owens, M. J., Crooker, N. U., & Schwadron, N. A. 2008, JGRA, 113, A11104
Pedregosa, F., Varoquaux, G., Gramfort, A., et al. 2011, J. Mach. Learn. Res., 12, 2825
Rosenbauer, H., Schwenn, R., Marsch, E., et al. 1977, JGZG, 42, 561
Saito, S., & Gary, S. P. 2007, JGRA, 112, A06116
Scudder, J. D., & Olbert, S. 1979, JGRA, 84, 6603
Stansby, D., Salem, C., Matteini, L., & Horbury, T. 2018, SoPh, 293, 155
Štverák, Š., Maksimovic, M., Trávníček, P. M., et al. 2009, JGRA, 114, A05104
Vocks, C., Salem, C., Lin, R. P., & Mann, G. 2005, ApJ, 627, 540
Whittlesey, P. L., Larson, D. E., Kasper, J. C., et al. 2020, ApJS, 246, 74
Woodham, L. D., Horbury, T. S., Matteini, L., et al. 2020, A&A, 650, L1
Zouganelis, I., Maksimovic, M., Meyer-Vernet, N., Lamy, H., & Issautier, K. 2004, ApJ, 606, 542